# The Effectiveness of Virtual R&D Teams in SMEs: Experiences of Malaysian SMEs

**Nader Ale Ebrahim**[†]
Department of Engineering Design and Manufacture, Faculty of Engineering,
University of Malaya Kuala Lumpur, Malaysia

**Salwa Hanim Abdul Rashid**
Centre for Product Design and Manufacturing, Faculty of Engineering,
University of Malaya, 50603, Kuala Lumpur, Malaysia

**Shamsuddin Ahmed**
Department of Engineering Design and Manufacture, Faculty of Engineering,
University of Malaya Kuala Lumpur, Malaysia

**Zahari Taha**
Faculty of Manufacturing Engineering and Management Technology,
University Malaysia Pahang, 26300 Gambang, Pahang, Malaysia

**Abstract.** The number of small and medium enterprises (SMEs), especially those involved with research and development (R&D) programs and employed virtual teams to create the greatest competitive advantage from limited labor are increasing. Global and localized virtual R&D teams are believed to have high potential for the growth of SMEs. Due to the fast-growing complexity of new products coupled with new emerging opportunities of virtual teams, a collaborative approach is believed to be the future trend. This research explores the effectiveness of virtuality in SMEs' virtual R&D teams. Online questionnaires were emailed to Malaysian manufacturing SMEs and 74 usable questionnaires were received, representing a 20.8 percent return rate. In order to avoid biases which may result from pre-suggested answers, a series of open-ended questions were retrieved from the experts. This study was focused on analyzing an open-ended question, whereby four main themes were extracted from the experts' recommendations regarding the effectiveness of virtual teams for the growth and performance of SMEs. The findings of this study would be useful to product design managers of SMEs in order to realize the key advantages and significance of virtual R&D teams during the new product development (NPD) process. This is turn,  leads to increased effectiveness in new product development's procedure.

## 1. INTRODUCTION

Small and medium-sized enterprises (SMEs) are major contributors for industrial economies (Eikebrokk and Olsen, 2007). The significance of SMEs in economic growth has rendered SMEs a central element in much recent policymaking (Hoffman *et al*., 1998). SMEs appear to be appropriate units as network nodes due to their lean structures, adaptability to market evolution, active involvement of versatile human resources, ability to establish subcontracting relations and good technological level of their products (Mezgar *et al*., 2000). SMEs possess advantages with regards to flexibility, reaction time and innovation capacity, and therefore SMEs play a major role in the new economy (Raymond and Croteau, 2006). Gassmann and Keupp (2007) found that managers of SMEs should invest less in tangible assets and more in areas which would directly enhance their future competitive advantage such as R&D, which would generate knowledge, as well as in their employees' creativity to stimulate incremental innovations in existing technologies. A crucial trend for enabling the creation and transfer of new knowledge in and to SMEs is by the development of virtual collaborative environments and networks to increase their innovation abilities as a single unit and capabilities of the network as a whole (Flores, 2006). Virtuality has been presented as a solution for SMEs aiming to increase their competitiveness (Pihkala et al., 1999). Virtual teams reduce time-to-market for new products (May and Carter, 2001). Lead time or time-to-market has been generally accepted as one of the vital keys for success in manufacturing companies (Sorli et al., 2006).

---

† : Corresponding Author

Ale Ebrahim et al. (2009a, 2010) derived the strengths and weaknesses of virtual teams in SMEs in their recent comprehensive reviews. The effectiveness of virtual teams in Malaysian manufacturing SMEs has not been reported, and therefore, the main objective of this study is to present the primary benefits of virtual teams for the growth of SMEs. The scope of this study is limited to the experiences of Malaysian manufacturing SMEs' expertise, which involve virtual teams. In this paper, the effectiveness is related to the performance and collaboration within virtual teams in order to reduce costs and time of R&D projects. This paper presents a portion of the results obtained from an empirical research carried out during the past two years within manufacturing SMEs in Malaysia. In moving towards virtual R&D teaming, an understanding of existing practices is important. In this paper, a review of recent literature pertaining to virtual R&D teams is presented, whereby the primary definition of virtual R&D teams and its relationship with SMEs are introduced. Following this, the research methodology and data analyses are detailed, and the directions for future research are presented in the final section of this paper.

## 2. VIRTUAL R&D TEAMS AND SMEs

Gassmann and Von Zedtwitz (2003) defined "virtual team as a group of people and sub-teams, which interact through interdependent tasks guided by common purpose and work across links strengthened by information, communication, and transport technologies." Another definition suggests that virtual teams are distributed work teams whose members are geographically dispersed and their works are coordinated mainly with electronic information and communication technologies (e-mail, video-conferencing, telephone, etc.) (Hertel et al., 2005). Among the different definitions of virtual teams, the following concept is one of the most widely accepted definitions (Ale Ebrahim et al., 2009c): "Virtual teams are small temporary groups of geographically, organizationally and/or time dispersed knowledge workers who coordinate their work, predominantly with electronic information and communication technologies in order to accomplish one or more organization tasks" (Ale Ebrahim et al., 2009b). Virtual R&D team is a form of a virtual team, which includes the features of virtual teams and concentrates on R&D activities. The members of a virtual R&D team utilize different degrees of communication technology to complete the research without space, time and organizational boundaries.

SMEs are not scaled-down versions of large companies as they possess different characteristics which distinguish them from large corporations. SMEs vary across different countries and cultures, and they are independent, multi-tasked and cash-limited as well as based on personal relationships and informality. Additionally, SMEs are managed actively by the owners, highly personalized, largely localized within their areas of operation and are largely dependent on internal sources for financial growth (Perrini et al., 2007). In order to survive in the global economy, SMEs have to improve their products and processes by exploiting their intellectual capital in a dynamic network of knowledge-intensive relations inside and outside their borders (Corso et al., 2003). Therefore, if small firms intend to create a step change in their technological and innovation base, they may have to rethink their approach to cooperation (Hanna and Walsh, 2002). SMEs need to cooperate with external partners to compensate for other competencies and resources. This is especially the case for R&D, in which SMEs face specific problems compared with large firms (Pullen et al., 2008). Levy et al. (2003) stated that SMEs are knowledge creators; however, they are poor in knowledge retention. They need to be proactive in knowledge sharing arrangements to recognize that knowledge has value, and the value added is derived from knowledge exchange (Egbu et al., 2005). Virtual R&D teams can provide such knowledge sharing. There is a general movement towards virtual R&D teams, as virtual R&D teams facilitate the spreading of risks and sharing or costs among a network of companies (Gassmann and Von Zedtwitz, 1999, Kratzer et al., 2005). Hence, virtual teams are important mechanisms for organizations such as SMEs seeking to leverage scarce resources across geographic and other boundaries (Munkvold and Zigurs, 2007).

## 3. METHODOLOGY

The data for this research was gathered from desk study and survey. Web-based questionnaires were designed and delivered to Malaysian manufacturing SMEs, which included close-ended and open-ended questions. This study clustered one open-ended question. Clustering involves searching the data for related categories with similar meaning. This analysis is known as Thematic Analysis since the main purpose during the start of the analysis is to look for themes. When a set of themes is formed, more advanced analyses can be employed to look for clusters and patterns among them (Abdul Rashid, 2009). In this analysis, any sentences which provide significant meaning were extracted and organized into different categories.

The Effectiveness of Virtual R&D Teams in SMEs                    111## 4. DATA COLLECTION AND ANALYSES

The research was targeted at manufacturing SMEs within Malaysia, which employed virtual teams in their organizations. Online questionnaires were sent to relevant SMEs in order to obtain the viewpoints from experts involved with virtual teams in SMEs. Denscombe (2006) encouraged social researchers to use web-based questionnaires with confidence, and therefore online questionnaires were distributed to SMEs in Malaysia via email. The participants were directed to a website, and the surveys were completed online.

The questionnaires consisted of three sections, as follows:

a) Demographic information: The results obtained from this section enable the selection of suitable enterprises which complied with the definition of SMEs.

b) Current status of virtual teams: The first question in this section clarified the utilization of virtual teams in the enterprises. Respondents who selected "No" in answer to the question indicate that the organizations did not possess experience with virtual teams, and were directed to Section C in the questionnaires. The final open-ended question which concerns the effectiveness of virtual teams on the organization's growth and performance, were analyzed in this research.

c) Requirements for establishing virtual teams: The results of this section was not included in this research.

The surveys were tested preliminarily among 12 experts, followed by improvements, modifications and distribution. Finally, questionnaires consisting of open and close-ended questions were distributed to 356 Malaysian manufacturing SMEs. The major target groups with regards to the size of the organization and industrial field were Managing Directors, R&D Managers, New Product Development Managers, Project and Design Managers as well as appropriate personnel who were involved significantly with R&D issues in the organizations. A total of 74 usable questionnaires were received, which represented a 20.8 percent return rate. The response rate was deemed satisfactory since accessing high-rank personnel was difficult. Table 1. It was found that a total of 42 SMEs fulfilled the criteria of this research and therefore the remaining respondents were dropped from the analysis. Descriptive statistics were used to analyze the responses. Table 2 shows the frequency of using virtual teams among the sampled Malaysian SMEs. The results showed that 33.3% SMEs employed virtual teams. This indicates that applications of virtual teams in manufacturing SMEs are still in its infancy.

**Table 1.** Summary of online survey data collection.

| Number of emails sent to Malaysian Firms | 2068 |
|---|---|
| Total Responses (Click the online web page) | 356 |
| Total Responses/Received questionnaire (%) | 17.2 |
| Total Completed | 74 |
| Total Completed/Received questionnaire (%) | 20.8 |

It is known that open-ended questions provide fewer prompts and impose the fewest limits. It is for these reasons open-ended questions evoke the most authentic possible responses from respondents (Bobrow, 1997). Open-ended questions are good for prompting a respondent's attitude or feelings, likes and dislikes, memory recalls, opinions, or to request for additional comments. However, open-ended questions are time-consuming and particularly difficult to answer. After considering all advantages and disadvantages, only a few open-ended questions were used in the online questionnaires. In this research, only one open-ended question was considered, which was: *Please explain the total effectiveness of virtual team system/tool on the company's growth and performance, before and after implementation?*

**Table 2.** Cross-tabulation between country and virtual teams.

|  | Using Virtual Team | | Total |
|---|---|---|---|
|  | Yes | NO |  |
| Count | 14 | 28 | 42 |
| % | 33.3% | 66.7% | 100.0% |

## 5. RESPONDENTS' COMMENTS

It was found that a great majority of the respondents answered the open-ended questions. Summarizing the results of open-ended questions was not simple due to the different levels of management and individuals, subjective wording and phrasing of the responses. However, several good comments were selected, and are shown as quotes in Table 3. The comments represent the actual experiences of the respondents, which are in accordance with (Ebrahim *et al.*, 2010, May and Carter, 2001, Bouchard and Cassivi, 2004). The virtual teams' managers were a good source to confirm the benefits of virtuality due to their experiences. Since open-ended questions provide a rather qualitative information, simple thematic analysis was particular suitable to extract information from such questions. In this research, simple thematic analysis was performed by conducting two levels of clustering analysis. Thematic analysis is commonly used by qualitative researchers and is usually recognized as a tool rather than a method (Abdul Rashid, 2009). In this analysis, the data were clustered into two levels, whereby lower level is Level 2, and higher level is Level 1. Level 1 was then identified as theme. Table 4 shows the clusters and theme generated from the simple thematic analysis. From this analysis, it was found that there are four main benefits of virtual team/tool on the growth and performance of enterprises. These benefits are: reduced R&D costs and time, more effective R&D, better output and increased coordination.

**Table 3.** Comments on the effectiveness of virtual teams for the company's growth and performance (Compare before and after implementation).

| Case No. | Respondents' comments |
|---|---|
| 1 | Cost saving, time saving, and great convenience. These will enhance the flow of the projects of a company and speed up the progress of our work. |
| 2 | Reduce time consumption |
| 3 | Time and cost are saved. |
| 4 | Since we have different manufacturing location around the world, our marketing department is located away from R&D, the virtual tools are the one that brings us closer and helps in decision making, faster product release and meeting customer satisfaction. |
| 5 | Virtual team system/tool is merely ASSISTANCE to the current workload. |
| 6 | Save time, money and energy |
| 7 | In my opinion, virtual team can make a good connection between the entire assets of organization. |
| 8 | With start virtual team system we improved in my performance |
| 9 | The virtual team system/tool is effective and can be helpful |
| 10 | In both it is seriously important. |
| 11 | 1) The company could growth faster, due to overcoming to distance and time by using virtual system<br>2) If system will be managed in an effective manner, the performance is increased due to power of the tools |
| 12 | We did some activities in our company to reduce costs as follows : 1-We arranged virtual network suppliers 2-They arranged R&D teams for our orders 3-our R&D department manage overall activities then we can reduced employees from 50 to less than 20 |
| 13 | 1) Capable for attracting experts and knowledge workers<br>2) declining ineffectual face to face meetings-improving work environment-Reducing time of trips |
| 14 | After correct implementation and good training of users, the growth of company is about 6 from 10 (10 is excellent and 0 is bad) |
| 15 | In my opinion it is impossible to work without such systems in the extremely mobile world we face these days. |
| 16 | Reduce unnecessary time waste and expedite product outcome |
| 17 | We demonstrate a positive annual trend in all factors important to us. |
| 18 | There is some effect but might be more effective while internal works are considered. In the case of international cooperation it depends strongly on consortiums formed for project executions |

## 6. CONCLUSIONS

Despite the enormous benefits of employing virtual R&D teams in manufacturing SMEs, the application of virtual teams by most enterprises is still in its infancy. The study showed that one-third of Malaysian manufacturing SMEs have employed virtual R&D teams. Competitive advantage is now becoming available to SMEs through geographically open boundaries created by virtual teams. Existing practices within Malaysian manufacturing SMEs experts, who were involved with virtual teams, proved four-fold benefiting from the cross-functional virtual R&D teams, namely: 1-Reduced R&D cost and time, 2-More effective R&D, 3-Better output, 4-Increased coordination. Virtual R&D teams give better team outputs, reduce time-to-market, reduce travel costs and demonstrate the ability to tap selectively into centers of excellence. Additionally, virtual R&D teams enable the use of the best talents regardless of location, giving a greater degree of freedom to individuals, shorter development times, and quicker response to changing business environments as well as higher team effectiveness and coordination. Therefore, the decision for setting up virtual R&D teams in SMEs is not a choice, but a necessity.

This paper is probably the first to present an empirical research on virtual R&D teams, which is limited to Malaysian manufacturing SMEs. Future research is needed to investigate the four-fold benefits of virtual R&D teams by a larger sample from different sectors. Although several studies have been carried out on the use of virtual R&D teams in large companies, applications within SMEs remain undocumented. Hence, future research should be focused on this gap and to search for a virtual collaborative system for SMEs which are dispersed geographically. Such a collaborative system should virtually link SMEs to enable the engaging members to focus on their specialized tasks as well as share their knowledge and experience (information resources). This will create agile manufacturing environments and enterprises.

**Table 4.** Clustered theme and cluster extracted from Table 3 (virtual team effectiveness).

| No. | Cluster Level 1 /Theme | Cluster Level 2 |
|---|---|---|
| 1 | Reduced R&D cost and time | Cost saving, Time saving<br>Reduce time consumption<br>Faster product release<br>Reduced employees<br>Reducing time of trips<br>Reduce unnecessary time wastage |
| 2 | More effective R&D | Speeds up work progress<br>Great convenience<br>Facilitates decision-making<br>Assists the current workload<br>Improved performance<br>Virtual team system/tool is effective<br>Capable of attracting experts and knowledge workers |
| 3 | Better output | Enhances the flow of projects of a company<br>Meets customer satisfaction<br>Increases performance<br>Improves work environment<br>Expedites product outcome<br>Demonstrates a positive annual trend |
| 4 | Increased coordination | Brings us closer<br>Good connection between the entire assets of organization |

## REFERENCES

Abdul Rashid, S. H. (2009), *An investigation into the material efficiency practices of UK manufacturers*, PhD Thesis, Cranfield University.

Ale Ebrahim, N., Ahmed, S., and Taha, Z. (2009a), Innovation and R&D Activities in Virtual Team, *European Journal of Scientific Research*, **34**, 297-307.

Ale Ebrahim, N., Ahmed, S., and Taha, Z. (2009b), Virtual R&D teams in small and medium enterprises: A literature review, *Scientific Research and Essay*, **4**, 1575-1590.

Ale Ebrahim, N., Ahmed, S., and Taha, Z. (2009c), Virtual Teams for New Product Develop-ment-An Innovative Experience for R&D Engineers, *European Journal of Educational Studies*, **1**, 109-123.

Ale Ebrahim, N., Ahmed, S., and Taha, Z. (2010), SMEs; Virtual research and development (R&D) teams and new product development: A literature review *International Journal of the Physical Sciences*, **5**, 916-930.

Bobrow, E. E. (1997), *The complete idiot's guide to new product development*, New York, Alpha Books.

Bouchard, L. and CASSIVI, L. (2004), Assessment of a Web-groupware technology for virtual teams, *IAMOT 2004*. Washington, D. C.

Corso, M., Martini, A., Paolucci, E., and PELLEGRINI, L. (2003), Knowledge management configurations in Italian small-to-medium enter-prises, *Integrated Manufacturing Systems*, **14**, 46-56.

Denscombe, M. (2006), Web-Based Questionnaires and the Mode Effect: An Evaluation Based on Completion Rates and Data Contents of Near-Identical Questionnaires Delivered in Different Modes, *Social Science Computer Review*, **24**, 246-254.

Ebrahim, N. A., Ahmed, S., and Taha, Z. (2010), Virtual R&D teams and SMEs growth: A comparative study between Iranian and Malaysian SMEs, *African Journal of Business Management*, **4**, 2368-2379.

Egbu, C. O., Hari, S., and Renukappa, S. H. (2005), Knowledge management for sustainable competitiveness in small and medium surveying practices, *Structural Survey*, **23**, 7-21.

Eikebrokk, T. R. and Olsen, D. H. (2007), An empirical investigation of competency factors affect-ing e-business success in European SMEs, *Information and Management*, **44**, 364-383.

Flores, M. (2006), IFIP International Federation for Information Processing. *Network-Centric Collaboration and Supporting Fireworks*, Boston: Springer.

Gassmann, O. and KEUPP, M. M. (2007), The competitive advantage of early and rapidly internationalising SMEs in the biotechnology industry: A knowledge-based view. *Journal of World Business*, **42**, 350-366.

Gassmann, O. and Von Zedtwitz, M. (1999), Organizing virtual R&D teams: towards a con-tingency approach. *In:* IEEE Management of Engineering and Technology, Technology and Innovation Management. PICMET '99, Portland International Conference on Management of Engineering and Technology, Portland, OR, USA. 198-199.

Gassmann, O. and Von Zedtwitz, M. (2003), Trends and determinants of managing virtual R&D teams, *R&D Management*, **33**, 243-262.

Hanna, V. and Walsh, K. (2002), Small Firm Networks: A Successful Approach to Innovation? *R&D Management*, **32**, 201-207.

Hertel, G. T., Geister, S., and Konradt, U. (2005), Managing virtual teams: A review of current empirical research, *Human Resource Management Review*, **15**, 69-95.

Hoffman, K., Parejo, M., Bessant, J., and Perren, L. (1998), Small firms, R&D, technology and innovation in the UK: a literature review, *Technovation*, **18**, 39-55.

Kratzer, J., Leenders, R., and Engelen, J. V. (2005), Keeping Virtual R&D Teams Creative, *Research Technology Management*, **1**, 13-16.

Levy, M., Loebbecke, C., and Powell, P. (2003), SMEs, coopetition and knowledge sharing: the role of information systems, *European Journal of Information Systems*, **12**, 3-17.

May, A. and Carter, C. (2001), A case study of virtual team working in the European automotive industry, *Inter-national Journal of Industrial Ergonomics*, **27**, 171-186.

Mezgar, I., Kovacs, G. L., and Paganelli, P. (2000), Cooperative production planning for small- and medium-sized enterprises, *International Journal of Production Economics*, **64,** 37-48.

Munkvold, B. E. and Zigurs, I. (2007), Process and technology challenges in swift-starting virtual teams, *Information and Management*, **44**, 287-299.

Perrini, F., Russo, A., and Tencati, A. (2007), CSR Strategies of SMEs and Large Firms. Evidence from Italy, *Journal of Business Ethics*, **74**, 285-300.

Pihkala, T., Varamaki, E., and Vesalainen, J. (1999), Virtual organization and the SMEs: a review and model development, *Entrepreneurship and Regional Development*, **11**, 335-349.

Pullen, A., Weerd-Nederhof, P. D., Groen, A., and Fisscher, O. (2008), Configurations of ex-ternal SME characteristics to explain differences in innovation performance, *High Technology Small Firms Conference* Twente University, Netherlands.

Raymond, L. and Croteau, A. M. (2006), Enabling the strategic development of SMEs through advanced manufacturing systems A configurational perspective, *Industrial Management and Data Systems*, **106**, 1012-1032.

Sorli, M., Stokic, D., Gorostiza, A., and Campos, A. (2006), Managing product/process knowledge in the concurrent/simultaneous enterprise environment, *Robotics and Computer-Integrated Manufacturing*, **22**, 399-408.